\documentclass[12pt,superscriptaddress]{revtex4}
\usepackage{graphicx}
\usepackage{dcolumn}
\usepackage{bm}
\usepackage{amssymb}
\usepackage{mathrsfs}

\begin{document}

\title{Radiation Pressure Dominate Regime of \\ Relativistic Ion Acceleration}

\author{T. Esirkepov}
 \altaffiliation[Also at ]{Moscow Institute of Physics and Technology, Dolgoprudnyi, Russia}
\affiliation{Kansai Research Establishment, JAERI, Umemidai 8-1 Kizu, Kyoto 619-0215, Japan}

\author{M. Borghesi}
\affiliation{The Queen's University of Belfast, Belfast BT7 1NN, UK}

\author{S. V. Bulanov}
 \altaffiliation[Also at ]{A. M. Prokhorov Institute of General Physics, RAS, Moscow, Russia}
\affiliation{Kansai Research Establishment, JAERI, Umemidai 8-1 Kizu, Kyoto 619-0215, Japan}

\author{G. Mourou}
\affiliation{Center for Ultrafast Optical Science, University of Michigan, Ann Arbor, Michigan 48109, USA}

\author{T. Tajima}
\affiliation{Kansai Research Establishment, JAERI, Umemidai 8-1 Kizu, Kyoto 619-0215, Japan}

\begin{abstract}
The electromagnetic radiation pressure becomes dominant in the interaction
of the ultra-intense electromagnetic wave with a solid material, thus the
wave energy can be transformed efficiently into the energy of ions
representing the material and the high density ultra-short relativistic
ion beam is generated. This regime can be seen even with present-day
technology, when an exawatt laser will be built. As an application,
we suggest the laser-driven heavy ion collider.
\end{abstract}



\maketitle

\section{Introduction}
The permanent  interest in the problems of the interaction of
relativistically strong electromagnetic radiation with matter finds broad
applications in the development of new concepts of charged particle
acceleration under space and laboratory conditions.  The generation of high
energy particles, both electrons and ions, when strong electromagnetic
radiation interacts with a plasma, is a well known basic phenomenon. However
in the limit of extremely high radiation intensity it acquires new features.
Under the earth conditions, an extremely strong electromagnetic wave can be
practically realized only with laser beams. When a multi-petawatt laser
pulse interacts with matter, conditions could be produced that were imagined
to occur only in astrophysical objects. This opens the way for experimental
studies of the properties of matter under these extreme conditions. With the
further increase of the laser power the laser based accelerators can provide
the beams of the ultrarelativistic ions, which makes feasible their
applications in the high-energy physics.

Direct laser acceleration of protons to relativistic energies requires
intensity
$I_p=4.6\times 10^{24}\,\mbox{W}/\mbox{cm}^2 \times (1\,\mu\mbox{m}/\lambda)^2$,
corresponding to the dimensionless amplitude
$a \equiv eE/m_e \omega c = m_p/m_e \approx 1836$,
where $E$, $\lambda$, and $\omega$
are the electric field, wavelength, and frequency of the electromagnetic (EM) wave,
$e$ and $m_e$ are the electron charge and mass,
and $m_p$ is the proton mass.
In a plasma, due to collective effects,
protons can gain relativistic energies at much less intensity, about 
$10^{21}\,\mbox{W}/\mbox{cm}^2 \times (1\,\mu\mbox{m}/\lambda)^2$,
as is exemplified in the theory of
the strongly nonlinear hybrid electron-ion wakefield
induced by a short EM wave packet with
the dimensionless amplitude $a$ greater than
$(m_p/m_e)^{1/2} \approx 43$
and
the Coulomb explosion of an overdense plasma region
with the size of a few microns
when a relativistically strong EM wave sweeps all the electrons away
\cite{1GeV}.
In general, the laser-driven ion acceleration
arises from charge separation caused by the EM wave.
Various regimes have been discussed in the framework of this concept:
the plasma thermal expansion into vacuum \cite{GPP-66};
the Coulomb explosion of a strongly ionized cluster \cite{Ditmier-96};
transverse explosion of a self-focusing channel \cite{Sarkisov-99};
ion acceleration in the strong charge separtion field
caused by a quasi-static magnetic field \cite{Kuznetsov-01}.

In this paper we present the regime of
the high density ultra-short {\it relativistic} ion beam generation
from a thin foil by an ultra-intense EM wave. 
We call this regime the `laser piston' (LP).
In contrast to previously discussed schemes,
in this regime
the ion beam generation is highly efficient
and the ion energy per nucleon is proportional to the laser pulse energy.
Our analytical estimation conforms to
the result of three-dimensional (3D) particle-in-cell (PIC) simulations.
In comparison with the experimental experience of present-day Petawatt lasers,
the LP regime
predicts yet another advantage
of the Exawatt lasers, in addition to possible applications
depicted in \cite{TM}.

\section{Radiation Pressure Dominant Regime}
We distinguish the following two stages of the LP operation.
(i)
A relativistically strong laser pulse irradiates a thin foil
with thickness $l$ and electron density $n_e$.
The laser pulse waist is sufficiently wide,
so the quasi-one-dimensional geometry is in effect.
Electrons are quickly accelerated up to $v_e\sim c$
by the transverse electric field, $E_L$, of the laser pulse
and they are pushed in the forward (longitudinal) direction by the
force $|e {\bf v}_e\times {\bf B}_L/c| \approx e E_L$, where $B_L$ is the
magnetic field of the laser pulse.
Assume that all the electrons are displaced in the longitudinal direction,
then the charge separation field,
$E_\parallel = 2\pi e n_e l < E_L $, between 
the electron and ion layers does not depend on the separation distance.
In this longitudinal field the ion energy
${\cal E}_i = (m_i^2 c^4 + (eE_\parallel c t)^2 )^{1/2}$
become relativistic in a time of the order of
$t_{ri} = (m_i c/e E_\parallel)$.
We find that the ion layer can be accelerated up to relativistic energies
during $N_L$ laser cycles
under the condition
$E_L > 2\pi e n_e l \gtrsim m_i \omega c/ 2\pi e N_L$.
Hence, to produce relativistic protons in one laser cycle
we need an EM wave with $E_L > 300 m_e\omega c/e$,
$I_L > 1.2\times 10^{23}\,\mbox{W}/\mbox{cm}^2 \times (1\,\mu\mbox{m}/\lambda)^2$,
and the pulse waist must be much greater than $c t_{ri}\simeq\lambda$.
(ii) 
The accelerated foil, which consists of the electron and ion layers,
can be regarded as a relativistic plasma mirror co-propagating with the laser pulse.
Assume that the laser pulse is perfectly reflected from this
mirror. In the laboratory reference frame,
before the reflection it has the energy ${\cal E}_L\propto E_L^2 L$,
after the reflection its energy becomes much lower:
$\widetilde{\cal E}_L\propto \widetilde{E}_L^2\widetilde{L} \approx
E_L^2 L/4\gamma^2$. Here the incidence laser pulse length is equal to
$L$, the reflected pulse length $\widetilde{L}$ is longer by factor of $4\gamma^2$,
and the transverse electric field
is smaller by factor $4\gamma^2$, where $\gamma=( 1 - v^2/c^2 )^{-1/2}$
is the Lorentz factor of the plasma mirror.
Hence the plasma mirror acquires the energy
$(1-1/4\gamma^2){\cal E}_L$ from the laser pulse.
At this stage the plasma (the electrons and, hence, ions)
is accelerated due to the radiation pressure.
The radiation momentum is transfered to ions
through the charge separation field
and the `longitudinal' kinetic energy of ions
is much greater than that of electrons.
We note that the specified intensity is close to the limit
where individual electrons undergo a substantial
radiation friction effect \cite{Rad-frict}.
However, when the foil is accelerated up to relativistic energy,
this effect becomes weaker because its strength measure, $4 \pi r_e/3 a^3 \lambda$,
is reduced by $2\gamma$ in the foil reference frame
($r_e=e^2/m_e c^2$ is the classical electron radius).

We notice a connection of the plasma layer acceleration scheme,
presented above, with a mechanism of the
ion acceleration proposed by V. I. Veksler \cite{Veksler},
as well as with the ``snow plow'' acceleration mechanism 
revealed in Ref. \cite{SnowPlow}.
Formulated in the mid 1950's Veksler's concept
of the  collective  acceleration of ions
in an electron-ion bunch  moving in a strong electromagnetic wave
had a great influence both upon
particle accelerator technology and plasma physics.
Up to now its direct realization
was considered at moderate
driving EM radiation intensities,
when transverse instabilities
impede the acceleration
\cite{Ask}.
As we show below using 3D PIC simulations,
in our scheme the transverse instabilities
are supressed or retarded due to the following.
(i)
The plasma layers become relativistic
quickly, during one or more laser wave periods,
in the first stage of the acceleration.
Due to relativistic effects the transverse instabilities
grow in the laboratory frame $\gamma$ times slower 
than in the plasma reference frame.
(ii)
The radiation pressure causes a stretching
of the plasma mirror in the transverse direction,
so the transverse instabilities can be retarded
similarly to the slowing down of the Jeans instability
in the theory of the expanding early universe
\cite{Jeans}.

\section{PIC Simulation}
\begin{figure}
\includegraphics{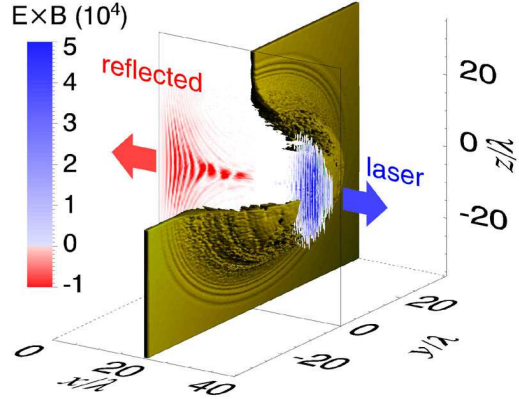}
\caption{The ion density isosurface for $n = 8 n_{cr}$
(a quarter removed to reveal the interior)
and the $x$-component of the normalized Poynting vector
$(e/m_e\omega c)^2\,{\bf E}\times{\bf B}$
in the $(x,y=0,z)$-plane
at $t = 40 \times 2\pi/\omega$.
}
\label{fig-1}
\end{figure}

\begin{figure}
\includegraphics{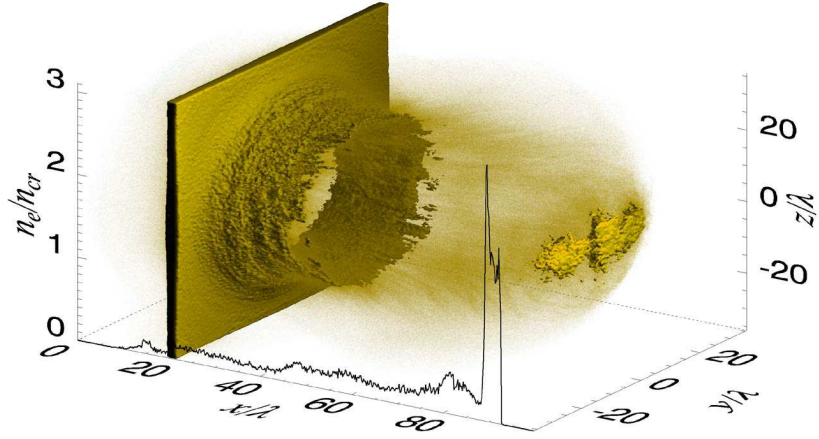}
\caption{The isosurface for $n = 2 n_{cr}$,
green gas for lower density
at $t = 100 \times 2\pi/\omega$;
the black curve shows the ion density
along the laser pulse axis.
}
\label{fig-2}
\end{figure}

\begin{figure}
\includegraphics{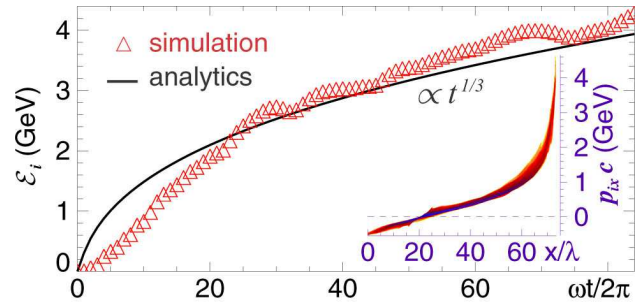}
\caption{The maximum ion kinetic energy versus time
and the ion phase space projection $(x,p_x)$ at $t = 80 \times 2\pi/\omega$.
}
\label{fig-3}
\end{figure}

\begin{figure}
\includegraphics{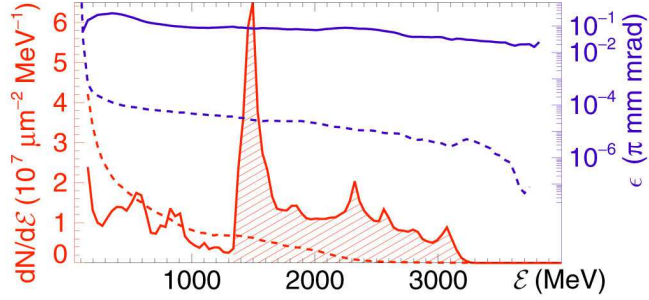}
\caption{The energy spectrum (red) and transverse emittance (blue) of
ions (solid) and electrons (dashed)
located in the box $50\lambda<x<80\lambda,-\lambda<y,z<\lambda$
at $t=80\times 2\pi/\omega$.
The hatched region containes $2.7\times10^{10}\mu$m$^{-2}$
particles per cross-section.
Values correspond to $\lambda=1\mu$m.
}
\label{fig-4}
\end{figure}

In order to examine the present scheme
in three-dimensional geometry,
whose effects may play a crucial role in
the dynamics and stability of the plasma layer
under the action of a relativistically strong laser pulse,
we carried out 3D PIC simulations
with the code {\sf REMP}
based on the ``density decomposition'' scheme \cite{REMP}.
In the simulations
the laser pulse is linearly polarized along the $z$-axis;
it propagates along the $x$-axis.
Its dimensionless amplitude is $a=316$ corresponding to the peak intensity
$I = 1.37\times 10^{23}\,\mbox{W}/\mbox{cm}^2 \times (1\,\mu\mbox{m}/\lambda)^2$.
The laser pulse is almost Gaussian with FWHM size
$8\lambda\times 25\lambda\times 25\lambda$,
it has a sharp front starting from $a=100$,
its energy is ${\cal E}_L = 10\,\mbox{kJ} \times (1\,\mu\mbox{m}/\lambda)^2$.
The target is a $1\lambda$ thick plasma slab with density
$n_e = 5.5\times 10^{22}\,\mbox{cm}^{-3} \times (1\,\mu\mbox{m}/\lambda)^2$,
which corresponds to the Langmuir frequency
$\omega_{pe} = 7\omega$.
For $\lambda\simeq 1\,\mu\mbox{m}$
the laser pulse electric field
is strong enough to strip even high-Z atoms
in much shorter time than the laser wave period,
thus we assume that the plasma is fully ionized.
The ions and electrons have the same absolute charge
and their mass ratio is $m_i/m_e=1836$.
The simulation box size is $100\lambda\times 72\lambda\times 72\lambda$
corresponding to the grid size $2500\times 1800\times 1800$,
so the mesh size is $0.04\lambda$.
The total number of quasi-particles is $4.37 \times 10^9$.
The boundary conditions are periodic along the $y$- and $z$-axis and
absorbing along the $x$-axis for both the EM radiation
and the quasi-particles.
Simulation results are shown in figures \ref{fig-1}--\ref{fig-4},
where the space and time units are the laser wavelength $\lambda$
and period $2\pi/\omega$.

Figure \ref{fig-1} shows the ion density
and $x$-component of the EM energy flux density (the Poynting vector).
We see that the region of the foil
corresponding to the size of the laser focal spot
is pushed forward.
Although the plasma in the foil is overcritical,
it is initially {\it transparent} for the laser pulse
due to the effect of relativistic transparency (see e. g. Ref. \cite{Rel-Transp}).
Therefore a portion of the laser pulse passes through the foil.
Eventually it accelerates electrons and,
as a result of the charge separation,
a longitudinal electric field is induced.
This can be interpreted as a rectification of the laser light,
by the analogy with a rectifier in electrical engineering:
the transverse oscillating electric field is
transfromed into a longitudinal quasistatic electric field.
The dimensionless amplitude of the longitudinal field is 
$a_\parallel \approx 150$ corresponding to
$E_\parallel = 4.8\times 10^{14}\,\mbox{V}/\mbox{m} \times (1\,\mu\mbox{m}/\lambda)$.
The typical distance of the charge separation is
comparable with the initial thickness of the foil
and is much less than the transverse size of the region being pushed.
The ion layer is accelerated by this longitudinal field.
This stage corresponds to the first step of the LP scheme.
As the foil moves more and more rapidly,
in its proper frame the incident wavelength increases,
thus the accelerating foil becomes less transparent with time.

As seen in the cross-section of the Poynting vector in Fig. \ref{fig-1},
the thickness of the red stripes,
corresponding to half of the radiation wave length,
increases from left to right (along the $x$-axis).
The increase is weaker at the periphery (in the transverse direction);
correspondingly, we see distinctive phase curves.
This `anisotropic red shift' results from the relativistic Doppler effect
when the laser pulse is reflected from the
co-propagating accelerating and deforming relativistic mirror.
The red shift testifies that
the laser pulse is expending its energy for
the acceleration of the plasma mirror,
as specified above in the second stage of the LP scenario.
The foil is transformed into a ``cocoon'' where
the laser pulse is almost confined.
The accelerated ions form a nearly flat thin {\it ``plate''}
with high density, Fig. \ref{fig-2}.

Figure \ref{fig-3} shows the ion maximum energy versus time
and the ion phase space plot.
The dependence is initially linear; at later times it scales as $t^{1/3}$.
The ion and electron energy spectra and transverse emittances
are presented in Fig. \ref{fig-4}.
The number of ions in the {\it plate}
is ${\cal N}_i=2\times 10^{12}$,
their energies are from $1.3$ GeV to $3.2$ GeV.
The efficiency of the energy transformation from laser to ions
in this range is 10\%; efficiency for all ions is 30\%.
The ion bunch density is $3\times 10^{21}$ cm$^{-3}$,
its duration is $20$ fs,
its transverse emmittance is less than $0.1\,\pi$ mm mrad.
As the {\it plate} is quasineutral,
the average longitudinal velocity of the electron bulk
is about that of ions, $v_{e\parallel}\approx v_{i\parallel}$,
thus the average longitudinal energy of electrons
is of the order of ${\cal E}_{e\parallel}\simeq (m_e/m_i){\cal E}_{i}$.
Correspondingly, the energy spectrum of electrons is peaked at much
less energy than that of ions.

According to our 2D and 3D simulations at lower intensities,
corresponding to Petawatt and multi-Petawatt pulses,
the interaction exhibits a continuous transition from
regimes of Refs. \cite{GPP-66, Sarkisov-99, Kuznetsov-01, 1GeV}
to the LP regime as the intesity increases.

\section{Ion Maximum Energy and Efficiency}
Here we estimate the ion maximum energy and the
acceleration efficiency in the model
of the flat foil driven by the EM radiation pressure,
as described above in the LP scenario.
In general, the radiation pressure on the foil
depends on its reflectance \cite{TeorPol}.
Assume that in the instantaneous reference frame,
where the foil is at rest,
the relative amplitudes of reflected and transmitted waves are
$\rho$ and $\tau$, respectively.
Here $|\rho|^2+|\tau|^2=1$ because of energy conservation.
The radiation pressure is the sum of
the incident, reflected and transmitted EM wave momentum fluxes,
$P=(E_L^{\prime 2}/4\pi)(1+|\rho|^2-|\tau|^2)=
(E_L^2/2\pi) (\omega'/\omega)^2 |\rho(\omega')|^2$.
Here the primed values correspond to the moving reference frame,
variables without a prime are taken in the laboratory frame.
In a quasi-one-dimensional geometry,
at the foil location $x(t)$
the laser electric field is $E_L = E_L(t-x(t)/c)$.
If the foil is accelerated as a single whole,
in its reference frame the incident radiation frequency becomes smaller
and smaller with time,
thus the reflection becomes more and more efficient and
$|\rho(\omega')|^2$ becomes closer to unity.
In fact, the foil reference frame
is not inertial since the foil is accelerated.
Hence, the EM wave frequency $\omega'$
decreases with time in this frame
as described in Ref. \cite{Hartemann}.
Nevertheless we can assume that
the acceleration is relatively small
and thus
$(\omega'/\omega)^2=(c-v)/(c+v)$, where
$v=dx/dt$ is the foil instantaneous velocity.

As is well known, the EM radiation pressure
is a relativistic invariant \cite{TeorPol},
therefore we can write the foil motion equation
as
\begin{equation}\label{eq-p}
\frac{dp}{dt} = \frac{E_L^2(t-x(t)/c)}{2\pi n_e l}
|\rho(\omega')|^2
\frac{\sqrt{ m_i^2 c^2 + p^2 } - p}{\sqrt{ m_i^2 c^2 + p^2 } + p}
\, ,
\end{equation}
where $p$ is the momentum of ions representing the foil.
In the simplest case, when $E_L=\mathrm{const}$ and $|\rho|^2=1$,
the solution $p(t)$ is an algebraic function of $t$.
For the initial condition $p=p_0$ at $t=0$
it can be written in a compact form as
$p = m_i c \left( \sinh (u) - {\rm csch}(u)/4 \right)$,
where $u=({1}/{3}) {\rm arcsinh}\left( \Omega t + h_0^3/2 + 3h_0/2 \right)$,
${\rm csch}(u)=1/\sinh (u)$,
$\Omega = 3E_L^2/{2\pi n_el m_i c}$
and $h_0 = p_0/m_i c + (1 + p_0^2/m_i^2 c^2)^{1/2}$.
The ion kinetic energy is
${\cal E}_{i\,{\rm kin}} = m_i c^2 \left( \sinh (u) + {\rm csch}(u)/4 - 1 \right)$.
As $t\rightarrow \infty$ it asymptotically tends to
${\cal E}_{i\,{\rm kin}}\approx m_i c^2 ( 3E_L^2 t / 8\pi n_el m_i c )^{1/3}$.
This motion is analogous to that of
a charged particle driven by a radiation pressure \cite{TeorPol},
but in our case the role of the Thomson cross-section
is played by the quantity $2/n_e l$.

To find an upper limit of the ion energy
acquired due to the interaction with a laser pulse
of finite duration,
we must include the dependence of the laser
EM field on space and time.
Because of the foil motion,
the interaction time can be much longer than the laser pulse duration $t_L$.
Therefore it is convenient to consider
the dynamics in terms of the dimensionless variable
\begin{equation}\label{eq-tau}
\psi = \int_{-\infty}^{t - x(t)/c}
\frac{E_L^2(\zeta)}{4\pi n_e l m_i c} d\zeta
\, ,
\end{equation}
which can be interpreted as the normalized
energy of the laser pulse portion that has been interacting
with the moving foil by time $t$.
Its maximum value is $\max\{\psi\} = {\cal E}_L/{\cal N}_i m_i c^2$,
where ${\cal E}_L$ is the laser pulse energy,
${\cal N}_i$ is the number of ions in the foil.
The solution of Eq. (\ref{eq-p}),
rewritten in terms of $\psi$,
gives the ion kinetic energy
\begin{equation}\label{eq-e-kin}
{\cal E}_{i\,{\rm kin}} \!=\!
m_i c^2 \frac{(2\varkappa\psi \!+\! h_0 \!-\! 1)^2}{2(2\varkappa\psi \!+\! h_0)}
\, ,
\;
\varkappa \!=\! \frac{1}{\psi}\int_0^{\psi} \!\! |\rho(\omega')|^2 d\psi
\, ,
\end{equation}
for the initial condition ${\cal E}_i(0)=(m_i^2 c^4 + p_0^2 c^2)^{1/2}$.
The upper limit of the ion kinetic energy and, correspondingly,
the laser to ion energy transformation efficiency
can be found from Eq. (\ref{eq-e-kin})
substituting $\psi=\max\{\psi\}$:
\begin{equation}\label{eq-e-limit}
\max\,\{ {\cal E}_{i\,{\rm kin}} \}
= \frac
{ 2\varkappa{\cal E}_L }
{ 2\varkappa{\cal E}_L + {\cal N}_i m_i c^2}
\;\frac{\varkappa{\cal E}_L}{{\cal N}_i}
\, ,
\end{equation}
where we set $p_0=0$ for simplicity. Here $\varkappa$
is the reflection coefficient, taken in the co-moving reference frame,
averaged over the foil motion path; $0<\varkappa\le 1$.
We see that if ${\cal E}_L \gg N_i m_i c^2/2$,
in this model
almost all the energy of the laser pulse
is transformed into ion kinetic energy.
Using Eq. (\ref{eq-e-limit}) and
the $t^{1/3}$ asymptotic dependence of the ion energy on time,
for given simulation parameters
we estimate the acceleration time and length as
$t_\mathrm{acc}\approx
(2/3)({\cal E}_L/{\cal N}_i m_i c^2)^2 t_L = 16$ ps
and
$x_\mathrm{acc}\approx c t_\mathrm{acc} =4.8$ mm,
respectively.
In the presented simulations
the acceleration time is limited by computer resources.
However, the analytical estimation Eq. (\ref{eq-e-limit})
allows us to conlude that at given simulation parameters
the limiting ion kinetic energy is 30 GeV.
Since the ion bunch is relativistic,
another ultra-intense laser pulse,
sent with proper delay,
can accelerate the bunch further.

\section{Conclusion}
In conclusion,
the LP regime of ultra-intense laser-plasma interaction
can be employed in a laser-driven heavy ion collider.
In the collision of two ion bunches
the number of reactions with cross-section $\sigma$
is $\mathscr{N} = \sigma {\cal N}_i^2 / s$,
where $s$ is the bunch sectional area.
Adopting the presented simulation parameters,
we obtain $\mathscr{N} \approx 2\times 10^{30} \sigma /\mbox{cm}^2$.
Provided that the energy is high enough
so that $\sigma\approx 10^{-24}$ cm$^2$,
we can get about a million events in a few femtosecond shot.
As suggested by two of co-authors (T. T. and G. M.) in Ref. \cite{TM},
one can get a short multi-Exawatt laser pulse with
sufficient contrast ratio ($10^{-12}$)
using the megajoule NIF facility and present-day technology.
Then the resulting ion bunch energy can be over $100$ GeV per nucleon,
which is suitable for the quark-gluon plasma studies \cite{Collider}.
The laser piston regime,
being one of the examples of what we call the Relativistic Engineering,
can give us a promising and unique tool for nuclear physics research.

\section*{Acknowledgments}
We thank M. Yamagiwa for significant remarks.
We thank
H. Daido,
J. Koga,
K. W. D. Ledingham,
P. Migliozzi,
K. Nishihara,
F. Pegoraro,
F. Terranova,
and A. V. Titov
for discussion,
APRC computer group for help.
This work is supported by MEXT, JST, INTAS 001-0233 and QUB/IRCEP.


\end{document}